\documentclass{article}

\usepackage{arxiv}

\usepackage[utf8]{inputenc} 
\usepackage[T1]{fontenc}    
\usepackage[english]{babel} 
\usepackage{graphicx}

\usepackage{hyperref}       
\usepackage{url}            
\usepackage{cleveref}       
\usepackage[style=authoryear,sorting=nyt,language=autobib]{biblatex}
\usepackage{doi}
\usepackage{csquotes}

\newcommand{\Rho}{\mathrm{P}}

\hypersetup{
	pdftitle={Contemporary Reaction to Gibbs's Statistical Mechanics}
	pdfsubject={physics.hist-ph, physics.class-ph},
	pdfauthor={Bruce D.~Popp},
	pdfkeywords={Gibbs, Elementary Principles in Statistical Mechanics, many-body dynamical system, ensemble, critical opalescence, Burbury, Hadamard, Lorentz, Orenstein},
}

\title{Contemporary Reaction to Gibbs's \emph{Statistical Mechanics}}
\date{December 14, 2024}
\author{ \href{https://orcid.org/0000-0001-9788-6466}{\includegraphics[scale=0.1]{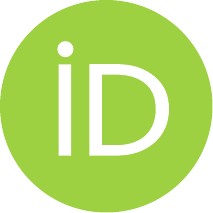}\hspace{1mm}Bruce D.~Popp} \\
	Independent Scholar \\
	Laguna Niguel, CA 92677 \\
	USA \\
	\texttt{BDPopp@Bien-Fait.com} \\
}

\begin{document}

\selectlanguage{english}

\maketitle

\begin{abstract}

J. Willard Gibbs published a book in 1902 on statistical mechanics that quickly received significant attention from his contemporaries because of the reputation that he had secured with his prior work on thermodynamics. People reading Gibbs's book were often familiar with Ludwig Boltzmann's work on the kinetic theory of gases. This article looks at the published response to Gibbs's book in the decade following its publication.

What did these readers get from reading Elementary Principles in Statistical Mechanics? How did they put that to use? Was it fruitful?

Samuel H. Burbury had been strongly critical of the assumptions on which Boltzmann's theory was built and had expressed a need for a theory without them. Burbury was pleased with the foundation and rigor of Gibbs's approach, since in his first three chapters he had built on analytical mechanics in the Hamiltonian formulation, a many-body dynamical system. 

Others found Gibbs hard to understand and some were critical. 

Hendrik A. Lorentz began teaching Gibbs's dynamical theory and supervised a thesis by Leonard Ornstein in which Orenstein developed applications built on grand canonical ensembles and correlations, which were contrary to Boltzmann's molecular disorder. Ornstein and Zernike expanded on material in that thesis to develop a theory of critical opalescence without infinities.
\end{abstract}


\section{ Introduction}

J. Willard Gibbs's book, \emph{Elementary Principles in Statistical Mechanics} \parencite{gibbs_elementary_1902}\footnote{Some wags noted that book's content was not elementary, as suggested by a misreading of the title; in some form, this was a common theme.}, was published well after most of Boltzmann's work on the kinetic theory of gases including his two-volume book, \emph{Lectures on Gas Theory}, were published. Contemporaries of Gibbs who read and reviewed \emph{Statistical Mechanics} were therefore familiar with Boltzmann's work. This set the stage for comparisons and objections.

What were those contemporary readers' responses---in book reviews, critical replies, and applications---to both Gibbs's content and to the distinct approach to the subject that Gibbs had taken?

This point has not received much attention. 

Jagdish Mehra \parencite*{mehra_josiah_1998} provides an admirable review of J. Willard Gibbs's work on statistical mechanics. Reaction to Gibbs's book plays a part in this and Mehra mentions Lorentz, Planck, Zermelo, and Ehrenfest and Ehrenfest-Afanassjewa. Mehra \parencite*[p.~1803]{mehra_josiah_1998} wrote, ``The book of Gibbs did not receive as much attention in the beginning as one might think today that it should have.'' This is probably true, but this paper considers more readers of Gibbs than Mehra mentions and looks more closely at their reactions. Mehra in the next sentence continues, ``The reason is that Gibbs wrote it in the same condensed, abstract and polished style which is typical of him, a style which arouses admiration more than it instills understanding in the reader at first contact.'' And that is both true and points to the importance of a second reading and study, and to the ultimate value of that effort.

As another example, João Príncipe \parencite*{principe_silva_reception_2008} does not distinguish the dynamical approach taken by Gibbs from the kinetic approach of Maxwell and Boltzmann and does not consider the close kinship of Poincaré's and Gibbs's independent development of many-body dynamical systems theory.

In this article, I aim to fill this need from the perspective of a physicist interested in what colleagues in physics and math from that era wrote about Gibbs's \emph{Elementary Principles in Statistical Mechanics}. The questions I asked myself when I read their writings were, ``What did they get from reading \emph{Statistical Mechanics}?'' and ``How did they put the physics to use?'' The answers range from ``very little'' to a theory of critical opalescence without infinities.

Before starting to review the response, the next section presents my contrast of the work of Boltzmann and Gibbs. I argue that Gibbs created a major shift in the physical basis by moving away from assumptions about velocities to analytical mechanics in the Hamiltonian formulation, and made a useful expansion of the ensemble approach by adding canonical ensembles and grand canonical ensembles with a focus on what was being averaged

The following section ({\S}~3) summarizes Gibbs's discussion (in Chapter~XII) of the density of coloring stirred in water. A section for this discussion is appropriate here because this chapter is considered difficult to understand and is singled out by several authors for specific discussion; this analogy with coloring is a subject of commentary and elaboration by Samuel Burbury and Henri Poincaré. 

The following section points out a translation into German, which appeared quickly, and a translation into French, which took much longer to appear. The translation into German was prepared by Ernst Zermelo and the translation into French was commissioned by Marcel Brillouin who also provided an introduction. He had earlier arranged for a translation into French of both volumes of Boltzmann's \emph{Lectures on Gas Theory} and provided introductions and notes for it \parencite{boltzmann_cons_1902,boltzmann_cons_1905}. 

The section on contemporaries' reactions covers writing from about 1904 to 1914. There, I survey writings by George H.~Bryan, Burbury (with a specific response to Chapter~XII), James H.~Jeans, Max Planck, Brillouin, Jacques Hadamard, Zermelo, Poincaré, Paul Ehrenfest and Tatiana Ehrenfest-Afanassjewa, Lorentz, Ornstein, Pierre Duhem, and Jan Kroo in chronological order.

The conclusion draws attention to notable points from the reactions.

\section{ Competing Physics}

Boltzmann's work was based on assumptions about the collisions and velocities of the atoms or molecules constituting the gas. While it is called a kinetic theory of gases, it could also be called a collisional theory. Referring to his theory as mechanical is therefore inaccurate, in my opinion.

Gibbs understood statistical mechanics as a many-body problem in analytical mechanics. Gibbs specifically used a Hamiltonian formulation of analytical mechanics. The Hamiltonian included the potential energy resulting from the forces between the bodies; he therefore relied on an implicit description of the forces encapsulated in the Hamiltonian function, instead of assumptions about the movement and collisions of atoms and molecules resulting from the action of the forces. In that way Gibbs overcame a criticism of Boltzmann's theory presented by Burbury. In presenting this criticism, Burbury said that Boltzmann's theory had neither  \textit{a priori} nor \textit{a posteriori} justification.

Gibbs avoided Boltzmann's assumptions and by using analytical mechanics started with a solid, \textit{a priori} foundation on which to build a structure as sound and reliable as his mathematical reasoning would allow. As Gibbs stated \parencite*[p.~x]{gibbs_elementary_1902} , ``Here, there can be no mistake in regard to the agreement of the hypotheses with the facts of nature, for nothing is assumed in that respect. The only error into which one can fall, is the want of agreement between the premises and the conclusions, and this, with care, one may hope, in the main, to avoid.'' As Poincaré \parencite{poincare_three-body_2017} had done 12 years earlier in his work on the equations of dynamics, much could be understood about the properties of the systems being described without explicit knowledge of the solutions of the equations or even, in Gibbs's presentation, an explicit mathematical statement of the potential energy---arising from, for example, the intermolecular forces---included in the Hamiltonian function.

I assert that Gibbs made a complete break from the work of his predecessor's kinetic (collisional) theory with his use of the Hamiltonian formulation of analytical mechanics. Yes, as, for example, indicated by Darrigol \parencite*[p.~17--18, 192--193]{darrigol_atoms_2021} , Henry W.~Watson had introduced the Hamiltonian with degrees of freedom, generalized coordinates and conjugate momenta. James Clerk Maxwell and Boltzmann picked up on this and in turn used it. The point is that Gibbs used that formalism in very different ways from Watson (and subsequently Maxwell and Boltzmann).

Specifically, Watson, in \parencite*{watson_treatise_1876} and again in a second edition in \parencite*{watson_treatise_1893}, starts his text on page~1 in both editions with the statement, ``A very great number of smooth and elastic spheres equal in every respect are in motion.'' That first paragraph continues by stating that the volume of the region of space is large compared to the total volume of the elastic spheres and that the mean time between collisions is infinitely large compared to the mean collision time. Watson has started by cleanly placing his book in the same context of elastic spheres with instantaneous collisions as Maxwell and Boltzmann. The Hamiltonian makes an anonymous cameo appearance on pages~12 and 13 of the first edition; in the second edition, it makes an appearance under its own name on pages 21--23 accompanied by an extensive footnote explaining that the reasoning of the demonstration in which it appeared in the first edition was ``fallacious.''

This is in no way comparable to the use that Gibbs made of the Hamiltonian in \emph{Statistical Mechanics}, where, as quoted above, it is introduced in the first sentence and is developed in the first three chapters. Gibbs's use of the Hamiltonian has its closest contemporary connection in the work of Poincaré on \emph{The Three-Body Problem and the Equations of Dynamics}. Both scholars make good use of the Hamiltonian formalism to determine the properties of the solutions without finding the solutions, or for that matter even defining the potentials going into the Hamiltonian.

A similar objection and reply involve ensembles used in determining statistical properties. Boltzmann introduced an ergode (but not the term) in 1871 in a paper referenced by Gibbs in his preface. There, Gibbs \parencite*[p.~viii]{gibbs_elementary_1902} wrote, ``The explicit consideration of a great number of systems and their distribution in phase, and of the permanence or alteration of this distribution in the course of time is perhaps first found in Boltzmann's paper.'' Some understand this as the origin of Gibbs's concept of ensembles. Gibbs defined canonical, then microcanonical and grand canonical ensembles. Boltzmann's ergode might correspond to a microcanonical ensemble. Stephen Brush \parencite[p.~507]{brush_kinetic_2003} wrote, ``While the word [ergode] thus originally pertained to what we now (following Gibbs) would call an \emph{ensemble}, it subsequently acquired a completely different meaning, primarily as a result of the influential article of the Ehrenfests which we will discuss below.'' Gibbs's development of ensembles therefore goes far beyond what Boltzmann might have meant in 1871 and has no connection with ergodic theory as understood later.

\subsection{ Boltzmann}

James Clerk Maxwell had written a series of articles on the theory of gases in the 1860s and '70s, and Ludwig Boltzmann had over a span of roughly 30 years from 1870 to 1900 worked on kinetic (collisional) theories of gases going through published increments, testing different directions, receiving comments and criticisms, followed by amplifications, amendments and even major changes in direction and fundamental approach. 

Much has been written about Boltzmann's work. His own book, \emph{Lectures on Gas Theory}, provides one entry point. Shortly after Boltzmann's death, a review article by Paul Ehrenfest and Tatiana Ehrenfest-Afanassjewa (\cite{ehrenfest_begriffliche_1911}, also available in translation as \cite{ehrenfest_conceptual_1990}), provided another synopsis, this time from their perspective and with their amplification. Much later, historical assessment and coverage was provided by Stephen Brush, Martin Klein, and Olivier Darrigol among others \parencite{brush_foundations_1967,klein_development_1973,darrigol_atoms_2021}. Instead of expanding the size of this article by summarizing these resources, the interested reader should consult them directly.\footnote{The reader is alerted that over the last century there have been ongoing developments to the theory bearing Boltzmann's name. One of the latest developments is referred to as ``the Boltzmann framework'' and described in section ``2.1 The Framework'' in \parencite{frigg_field_2008}. There, footnote~6 on page~9 cites its origin in work by Joel Lebowitz and Sheldon Goldstein. This appears to be an effort to stitch Boltzmann's name, a restriction to only microcanonical ensembles, and his formula for entropy onto a Hamiltonian dynamical systems approach. This Boltzmann framework was not part of the context in which Gibbs's book was published.} 

Boltzmann (and Maxwell before him) had assumptions about velocity fields, elastic collisions, uncorrelated collisions and pairwise interactions between hard spheres composing a gas. One statement of assumptions appears in the first paragraph of Watson's book, the first sentence of which is quoted above. Another is Boltzmann's assumption of molecular disorder (in German \emph{molekular-ungeordnet}) prevailing in the gas. (He uses molecular disorder to justify the assumption that velocities of collisions are uncorrelated.) The term introduced later by Paul Ehrenfest and Tatiana Ehrenfest Afanassjewa \parencite*{ehrenfest_begriffliche_1911} for these assumptions is \emph{Stosszahlansatz}. 

Burbury\footnote{Henry Watson and Samuel Burbury were lifelong friends from their meeting at Cambridge University. Burbury made contributions to Watson's book discussed above; they co-authored a book on generalized coordinates and another on electricity; and shared an interest in kinetic theory. \parencite{bryan_s_1911} }, an English barrister with a significant avocational interest in mathematics and physics, and notably kinetic theory of gases, published a book, \emph{A Treatise on the Kinetic Theory of Gases}, in 1899. The title is misleading since the book does not provide a general coverage of the theory. Indeed, it is a firm but deferential critique of the assumption that the velocities of the colliding elastic spheres are uncorrelated. Burbury referred to the state defined by the assumption that the velocities are uncorrelated as ``Condition A.'' Burbury \parencite*[p.~vi]{burbury_treatise_1899} states, ``It is no light thing to question a conclusion maintained by Boltzmann, if indeed he does maintain this conclusion for all substances, or for all gases irrespective of density. I can but state the objections to this theorem, and to a certain aspect of the H theorem, as they appear to me.'' Burbury asserts that Boltzmann's assumption that there is no correlation restricts the application of his theory to rarefied gases. Burbury follows this with constructive development of a theory for dense gases, even ones near liquefaction, that allows correlation of velocities through a parameter $b$.

While Burbury questions the validity of the assumption that the velocities are uncorrelated, he does, in several places, provide a clear statement of what this means. The assumption of molecular disorder was criticized by Burbury, Poincaré and other French probability theorists as not having a definition. It may even be a matter of ``Boltzmann knows it when he sees it.'' 

More fundamentally, Boltzmann's assumptions \textit{per se} are not confirmed or established as reliable \textit{a priori}. They consequently need some other confirmation. At best, this means subsequent confirmation, for example experimental, of the results obtained using the assumptions. In this sense Boltzmann's kinetic theory is built on ``an insecure foundation,'' and Gibbs's work, which carefully avoided assumptions, is based on analytical mechanics and is therefore well established and reliable.

\subsection{ Gibbs}

As can be seen in different places, there has been a significant difference in historical or philosophical interest in the work of Gibbs as compared to Boltzmann. For readers interested in more information about Gibbs's writing on statistical mechanics, it is worth consulting the articles in the Proceedings of the Gibbs Symposium (with specific attention to \parencite{klein_physics_1990,wightman_prescience_1990}. 

Gibbs started the first chapter of his book writing, ``We shall use Hamilton's form of the equations of motion for a system of $n$ degrees of freedom.'' And adds in a footnote ``We shall find that the fundamental notions of statistical mechanics are most clearly defined, and are expressed in the simplest form, when the momenta with the coordinates are used to describe the state of a system.'' Gibbs presents various tools of dynamical systems, such as (in his terms) extension-in-phase, conservation of extension-in-phase (Liouville's theorem), probability of phase, and Poincaré's recurrence theorem. This development of the theory of many-body Hamiltonian dynamical systems represents perhaps a substantial first third of this book.\footnote{The presence of this important foundation that Gibbs developed may surprise people who share the common perception that Gibbs's book only dealt with statistics of canonical ensembles and an alternative formula for entropy.} 

Because of his development of the many-body dynamics, Gibbs can avoid making any reference to the nature or properties of the bodies having the degrees of freedom. It is not a matter of delaying specifying the form of the Hamiltonian; much can be learned about the properties and statistics of the solutions without ever having the solutions or the specific Hamiltonian. This allows Gibbs to satisfy the objective that he set out in the preface, of avoiding ``building on an insecure foundation, [by resting] his work on hypotheses concerning the constitution of matter'' that do not correctly agree on the number of degrees of freedom of a diatomic gas. The Hamiltonian dynamical system approach had the advantage for Gibbs of eliminating any assumptions about the structures of the bodies, the number of degrees of freedom and the nature of the collisions. 

Poincaré, over 10 years earlier, had also introduced a many-body, dynamical systems approach \parencite{poincare_three-body_2017}. In this sense, these two books by Poincaré and Gibbs can be seen as companions that start by covering the same foundations of dynamical systems theory. Poincaré continued by extending his studies to celestial mechanics, nonlinear oscillators, and the properties of the solutions (for example stability) that can be inferred without any actual solution. In continuing work, Poincaré stayed with celestial mechanics and expanded to other examples of many body systems. In that context, we can understand why Poincaré read Gibbs's work with significant interest and appreciation as discussed below in {\S}~5.8.

Gibbs looked at the aggregate behavior of these systems of bodies. Gibbs in this book develops the statistics as phase space averages over an ensemble of systems. Notably he expanded the concept of ensembles. Gibbs defined three ensembles. First, he defined the canonical ensemble, then he defined the microcanonical ensemble and grand canonical ensemble in terms of the canonical ensemble. (Gibbs sometimes referred to the canonical ensemble as a petit canonical ensemble to distinguish it from the grand canonical ensemble.) 

Gibbs, in the beginning of Chapter X \parencite*[p.~115--117]{gibbs_elementary_1902} , defined \emph{microcanonical ensembles} in terms of a system in statistical equilibrium where the energy is bounded between an upper and lower bound, in the limit where the difference between the upper- and lower-bound approaches zero. Gibbs notes (on the following page) that we can regard ``the canonical ensemble as consisting of an infinity of microcanonical ensembles.''

Gibbs, in Chapter XV, introduces the concept of indistinguishable particles and the possibility that various kinds of particles can pass from one to the other. Gibbs \parencite*[p.~187--9]{gibbs_elementary_1902} first accepts that if two phases differ only because indistinguishable particles have swapped places than the phases are identical and he refers to them as \emph{generic phases}. Gibbs allows for the case where the exchange of similar particles is regarded as changing the phase and refers to it as a \emph{specific phase}. He notes that this definition means that there may be statistical equilibrium for generic phases, but without equilibrium for specific phases, and not vice versa. One generic phase may comprise a very large number of specific phases. Gibbs \parencite*[p.~189--91]{gibbs_elementary_1902} then considers ensembles composed of multiple kinds of particles; these are grand canonical ensembles. An ensemble composed only of particles of a single kind is petit canonical ensemble, and is the same as the canonical ensemble he defined in Chapter IV. A grand canonical ensemble is then a sum of petit canonical ensembles. Notably, the number of particles in the petit canonical ensembles may change.

The statistics of the Gibbs's ensembles involve averages over many copies of the phase space. For some people, this has been an obstacle to understanding averages and statistics. However, Poincaré, for example, appears to have had a clear understanding \parencite{poincare_reflexions_1906}. As discussed below, Poincaré and Orenstein both made effective use of grand canonical ensembles.

The sources for Gibbs's work are not known. In his preface to \emph{Statistical Mechanics}, Gibbs, as already discussed, mentions an 1871 article by Boltzmann on Jacobi's last multiplier as influential; Gibbs who does not indicate the actual influence of this article, but it is generally assumed to be its discussion of ensembles. It is also known that Gibbs had studied a late edition of Lagrange's \emph{Analytical Mechanics.} Gibbs does not mention either author in the body of \emph{Statistical Mechanics}. Given the nature of Gibbs's work habits and the long gestation of this book (discussed next), the work in his book was thoroughly his own, carefully considered and presented in the way he thought best.

We therefore have no indication how Gibbs might have understood a comparison of his work with that of his predecessors.

This single book by Gibbs on statistical mechanics from 1902 has an enduring note of finality.

Part of the finality of \emph{Elementary Principles in Statistical Mechanics} is because Gibbs died soon after its publication precluding any revised or expanded second edition. There could be no revised chapter on the approach to equilibrium, for example. Although Burbury addressed a question to Gibbs, as discussed below, Henry A. Bumstead replied posthumously for Gibbs. Even if, hypothetically, he lived his full life expectancy, it appears that Gibbs might not have returned to the subject in writing. Gibbs had over 25 years to return to his earlier work \emph{On the Equilibrium of Heterogeneous Substances} and did not prepare an additional work, whether a revised edition, supplement, or commentary. Although, Lynde Phelps Wheeler, in his excellent biography \emph{Josiah Willard Gibbs} \parencite*[p.~104]{wheeler_josiah_1951} indicates that Gibbs was discussing supplementary chapters with a publisher.

Another factor is the care and effort that Gibbs put into preparing his subjects for publication. He considered different perspectives on the subject and then chose the one best for development and presentation. A general indication of Gibbs's approach is given by Edwin Wilson, who noted \parencite[p.~546]{wilson_last_1961}, ``His publications are notable for their finished perfection; he had a keen sense of what was worth publishing and what was not.'' Referring to \emph{Elementary Principles in Statistical Mechanics}, Wheeler \parencite*[p.~154]{wheeler_josiah_1951} sees evidence, in a paper Gibbs presented in 1884 and in later notes for lectures for a course on dynamics and thermodynamics, that Gibbs was trying out ideas and methods for presenting the subject. 

This was confirmed with further detail by Martin Klein. He \parencite*[p.~12--13]{klein_physics_1990} states that the 1884 paper, ``On the Fundamental Formula of Statistical Mechanics with Applications to Astronomy and Thermodynamics,'' was presented to the American Association for the Advancement of Science. He also states that the fundamental formula mentioned in the title is Liouville's theorem, although Gibbs did not use that name. This title is also apparently the first use of the term statistical mechanics. Klein \parencite*[p.~14 and note 65]{klein_physics_1990} indicates that the lecture notes were from 1894 to 1895 and kept by George P. Starkweather. 

Gibbs's \emph{Elementary Principles in Statistical Mechanics} therefore had analytical mechanics ensembles and statistics, and had no assumptions, no gas, and no collisions.

\section{ Gibbs's Analogy to Coloring in Water}

A separate discussion of this analogy is important for several reasons. Henri Poincaré earlier presented a version of this analogy involving mixing of a barley grain in a heap of wheat and repeated Gibbs's version substituting wine for coloring; it was important to both of them as an example of mixing and irreversibility. It also attracted the interest of several reviewers because of the importance of understanding irreversibility. The use of an analogy here is also a part that makes this chapter very different from the first chapter and that attracted Burbury's notice. The analogy was also seen as difficult to understand and an effort to explain it may be appreciated.

Gibbs \parencite*{gibbs_elementary_1902} started Chapter~XII with a statement of the Poincaré recurrence theorem.\footnote{Gibbs does not reference Poincaré as a source of this theorem. It is conceivable that Gibbs developed an independent proof of the theorem. Further, since Gibbs's work underwent a long preparation as discussed above, it is possible that Gibbs had the recurrence theorem first but published later. } The proof that Gibbs provided relies on his principle of conservation of extension-in-phase which is the same as Liouville's theorem that Poincaré used in his proof of the recurrence theorem. Gibbs followed this proof and discussion with a question \parencite[p.~143]{gibbs_elementary_1902}, ``Let us next consider whether an ensemble of isolated systems has any tendency in the course of time towards a state of statistical equilibrium.'' To answer this, Gibbs showed---as Poincaré \parencite*{poincare_sur_1889} had sought to do and as Olsen \parencite*{olsen_classical_1993} subsequently showed rigorously---that in a Hamiltonian dynamical system the functions of phase are constant in time and that therefore using such a function ``We find{\dots} no approach towards statistical equilibrium in the course of time.'' We now see Gibbs stating, as Poincaré and Zermelo had before him, that for a dynamical system, there is no function of the phase that could correspond to an increasing entropy.

This shows us the importance to Gibbs of the issue that also attracted Poincaré's attention: how can a deterministic, analytical mechanical theory account for irreversibility and monotonically increasing entropy in a many-body system? This issue in Chapter~XII also attracted the attention of Gibbs's readers and continues to hold ours.

Gibbs \parencite*[p.~145]{gibbs_elementary_1902} looked at the problem of mixing coloring matter with water in this context of understanding this issue of irreversibility. He considered the average square of the density of the coloring matter in the water and noted that if the coloring matter started with a nonuniform density, then, subject only to the hydrodynamic condition of incompressibility, the density at any single point will remain unchanged. Yet, we have all seen that stirring the liquid leads to a uniform mixture; we observe an irreversible process.

Gibbs traces this contradiction to the concept of the density of the coloring matter and the volume for which it is measured. When tracking smaller and smaller elements of volume, as the stirring continues, the distribution of average square density does not change and the density in the elements does not become uniform. In contrast, if the volume and position of the elements are taken as fixed, the distribution of the average square density does change since the density of the elements becomes more uniform as the stirring continues. Gibbs \parencite*[p.~145--6]{gibbs_elementary_1902} stated that the case is ``Evidently one of those in which the limit of a limit has different values, according to the order in which we apply the processes of taking the limit.'' (The limits being smaller elements of volume and longer stirring time.)

\begin{figure}
    \centering
    \includegraphics[width=0.5\linewidth]{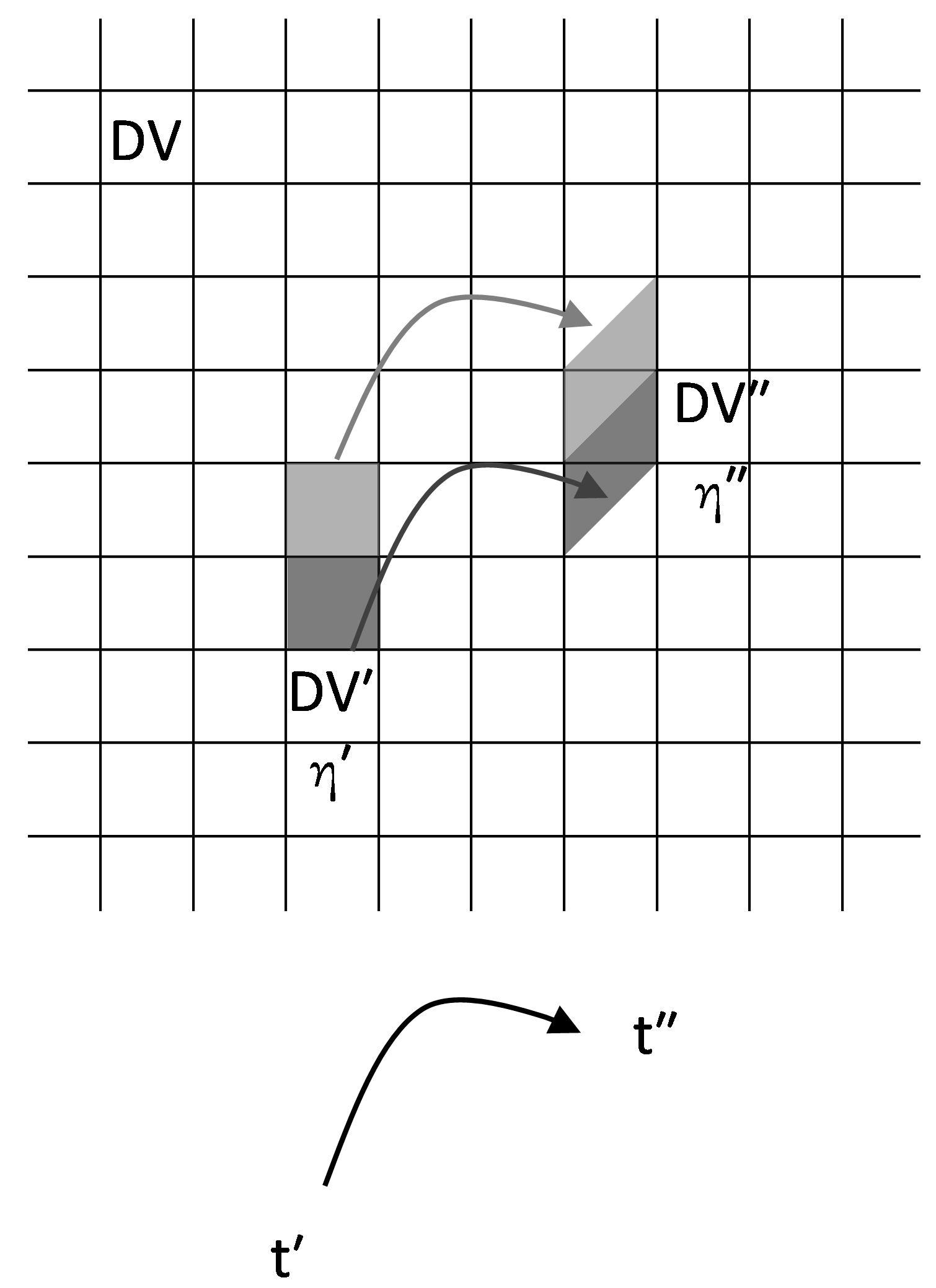}
    \caption{Movement of systems in phase space based on an example from Gibbs}
    \label{fig:traj__in_phase}
\end{figure}

After a second example involving mixing, Gibbs \parencite*[p.~148--51]{gibbs_elementary_1902} considered systems in phase space; that space has been divided into equal, small but not infinitesimal volumes, $DV$. For illustration, a sketch is provided in Figure \ref{fig:traj__in_phase}. He considers a system (here dark gray) that starts in one of the small volumes at time $t'$, with probability function $\eta '$. At some later time $t''$, the phase space volume of the system will be exactly the same and the system will be in a second small volume $DV''$, in part, and also in other small volumes, if the elapsed time ($t''-t'$) was sufficiently long. Other systems, starting in a different small volume (light gray) will also occupy the second small volume $DV''$, in part. The probability function in this second small volume at time $t''$ will then be $\eta ''$.

Gibbs states that for his purposes, it is the value of $\eta ''$ which is important. Along the orbit in phase space followed by the dark gray system, the probability function will be constant and in that part of the small volume~$DV ''$, the probability function will be nearly $\eta '$. In a situation of statistical equilibrium, the light-gray system will have nearly the same probability function and follow nearly the same orbit as the dark gray system, with the consequence that the probability function in $DV ''$ will also be nearly the same. If this is not a case of statistical equilibrium, then the two orbits may still be nearly the same, but no statement about the probability functions in $DV ''$ is possible.

This means that in the case of near statistical equilibrium the probability function in the small volume~$DV ''$ is nearly the average of the probability functions of the systems that followed nearly the same phase space orbit into the small volume. In the near-equilibrium situation, the coarse probability function coming from the rough average of the contributing probability functions is nearly accurate. But if the situation is far from statistical equilibrium, then the phase space orbits bringing systems into the small volume $DV ''$ may be different and the coarse probability function may be far from accurate.

Applying the analogy with the coloring matter makes it obvious that our ability to follow the orbits and the volumes $DV$ to smaller sizes is limited. The nature and consequences of this limitation were taken up by Burbury and Poincaré and are discussed below.

\section{ Translations}

Two translations of \emph{Elementary Principles in Statistical Mechanics}, into German and French, were prepared after contrasting delays. 

Ernst Zermelo prepared the translation into German \parencite{zermelo_elementare_1905}. It is not clear whether he prepared the translation at the urging of Max Planck or on his own initiative because of his interest in thoroughly studying the content. In his introduction, Zermelo \parencite[Editor's Forward]{zermelo_elementare_1905} indicated that he had two reservations; these were presented separately in \parencite{zermelo_notizen_1906} and are discussed in a later section. Apart from these reservations, which need to be taken seriously, Zermelo responded positively to the rigor and substantial interest he found in this work by Gibbs. 

By placing this emphasis on rigor in Gibbs's approach, Zermelo could have been suggesting that rigor was lacking in some of Boltzmann's work. A concern about Boltzmann's rigor was present in Zermelo's critique of Boltzmann in 1896. Poincaré also reacted to the lack of rigor in Boltzmann's concept of molecular disorder. Rigor is an area where there is a strong contrast between the work of Gibbs and Boltzmann.

In the concluding paragraph of his introduction, Zermelo marked his interest by stating his intent to share it, ``I hope that this work in its new form helps promote the interest in and understanding of statistical-mechanical problems among German physicists and mathematicians.''\footnote{Both quotes from Zermelo \parencite[Editor's Forward]{zermelo_elementare_1905} in this section were translated by Carola F.~Berger, October~29, 2022.} 

With my translator hat on, let me point out Zermelo's stated intent, his theory of translation. ``The translation is as close to the original as possible as the differences between the languages allow. The terminology, which mostly conforms to the international scientific language, was adopted as much as possible as well.'' This objective is certainly familiar to me and my translation colleagues.

Zermelo's translation of Gibbs's book into German appeared about three years after the book itself; the translation into French appeared a further 20 years later. Marcel Brillouin arranged for the translation of Gibbs's book into French; he had previously arranged for the translation of Boltzmann's \emph{Lectures on Gas Theory} into French \parencite{boltzmann_cons_1902,boltzmann_cons_1905}. In the introduction by Marcel Brillouin \parencite[Introduction]{gibbs_principes_1998} , he explained the delay as due to the death of the first translator and the late completion by another translator. Brillouin indicated that the interest in the book, and its translation, after a long delay was still strong and attributes that to the exceptional interest of the book and the powerful interest that it represents. As Zermelo did in his introduction, Brillouin compared Gibbs's book to Boltzmann's second volume of \emph{Lectures on Gas Theory} and lauds the better organization, presentation and clarity entirely due to Gibbs.

\section{ Reaction}

At the 1904 St. Louis Worlds' Fair, Poincaré and Boltzmann mentioned Gibbs's book in their addresses at the associated conference. As noted in the relevant section below, Poincaré stated the book was a little difficult to understand. 

Gibbs's book was also noticed by Einstein, who made an emphatic endorsement \parencite[p.250; Doc. 10]{einstein_collected_1993} \footnote{Doc. 10 from this collection was originally published in \emph{Annalen der Physik} vol.~34, 1911. For a discussion of the context of this note by Einstein see \parencite[\S~6.1]{navarro_gibbs_1998}. }:

\begin{displayquote}
I only wish to add that the road taken by Gibbs in his book, which consists in one's starting directly from the canonical ensembles, is, in my opinion, preferable to the road I took. Had I been familiar with Gibbs's book at the time, I would not have published those papers [in \emph{Annalen der Physik} in 1902 and 1903] at all, but would have limited myself to the discussion of just a few points. 
\end{displayquote}

In addition to these noteworthy mentions, there was a considerable response in print and this section reviews that response.

\subsection{ Bryan: A Very Fast Review}

Gibbs's book \parencite*{gibbs_elementary_1902} was published in March. In July of the same year, a review of this book written by George H.~Bryan was published in Nature \parencite{bryan_elementary_1902}. Bryan is known for his 1891 report to the British Association for the Advancement of Science on the connection between the second law of thermodynamics and dynamical principles. Even considering the short length of the review, the speed of its appearance is astonishing. 

Bryan \parencite{bryan_elementary_1902} starts with a discussion of the distinction of deductive and experimental side of science. He also notes a connection in spirit to the book by Watson, \emph{Kinetic Theory of Gases}. Watson indeed considered gases with more than three degrees of freedom and did so with generalized coordinates corresponding to each of those degrees.

Bryan remarked on two further, relevant points: the form of Gibbs's presentation ``can be studied quite independently of any molecular hypothesis as a purely mathematical deduction from the fundamental principles of dynamics'' and ``Prof.~Gibbs's work is not very easy to read, and it hardly seems appropriate to apply the title `elementary' to it.''

\subsection{ Burbury: A Sharp Critique of Boltzmann, Focus on Entropy}

We met Samuel H. Burbury in connection with his book published in 1899 with a critique of Boltzmann's assumption that atomic velocities are uncorrelated. Recall that three years before Gibbs's book appeared, Burbury wrote a book on kinetic theory as a vehicle for a critique of Boltzmann's assumptions. In that book, \emph{Kinetic Theory of Gases} \parencite*{burbury_treatise_1899}, Burbury described a \emph{fundamental assumption}. Assumption A or condition A corresponded to Boltzmann's assumption that atomic velocities are uncorrelated before a collision, which Boltzmann justified through his concept of molecular disorder. Assumption B was the opposite, it made no assumption about the presence of correlations of atomic motions.

Burbury brought a discussion of this fundamental assumption to his critique \parencite{burbury_mr_1903} of an article by James H.~Jeans on the kinetic theory of gases. That article was likely a mile marker on Jeans's path to his textbook, \emph{The Dynamical Theory of Gases} \parencite{jeans_dynamical_1904}. In the critique, Burbury again criticizes \emph{molekular ungeordnet}, ``But this has never been explained either by its author or by anybody else.'' And links it to Assumption A. Burbury goes on to indicate that he thinks that Jeans has correctly shown that Assumption A is mathematically impossible ``if the state of the system at any instant is a necessary consequence of its past history.'' And more Burbury indicates Jeans has shown that, ``Assumption A is not \textit{a priori} consistent{\dots} is not justified \textit{a posteriori}.'' So far, all of this adds to Burbury's established critique of Boltzmann. Burbury then states that, despite Jeans having shown this about Assumption A, Jeans continued his article about dynamical theory of gases on the basis of Boltzmann's orthodoxy, including Assumption A.~Jeans's response in print and his subsequent textbook suggest uncertainty or confusion about how to make use of this critique.

Gibbs's book came after Burbury's presentation of the fundamental assumption. It can be seen after the fact that Gibbs made no assumption about correlation of velocities (or molecular disorder) so it is an example of Assumption B. Presumably Burbury recognized this; it might also be why he did not present a general review of Gibbs's book and instead focused on the treatment of irreversibility.

In his reaction to Gibbs, the subject in this subsection, Burbury wrote about issues from Gibbs's Chapter~XII. This writing appeared in two papers \parencite{burbury_variation_1903,burbury_theory_1904}. Following the first paper, Henry A. Bumstead\footnote{That this response fell to Bumstead, following the death of Gibbs, seems to be related both to his coursework and involvement with Gibbs \parencite[p.~106--107]{page_henry_1929} and to his emerging connections with England and with English physicists \parencite{thomson_prof_1921}. Later in the year in which the reply was published, Bumstead left New Haven to spend a year in Cambridge at the Cavendish Laboratories and from 1917 to 1919 he was Scientific Attaché to the US Embassy in London. }, a professor at Yale University, published a response attempting to posthumously present an assessment of Gibbs's attitude to key issues raised by Burbury in the first paper.

Burbury, in his paper \parencite*{burbury_variation_1903} on Gibbs's \emph{Statistical Mechanics}, stated that his objective was to consider certain topics in Chapter~XII and prepared for that by reviewing the discussion and definition of density-in-phase in Chapter I. He (p.~252) indicates that the presentation by Gibbs of the Hamiltonian dynamical system approach in that chapter was easy ``owing to the extreme simplicity and elegance'' of its presentation by Gibbs. He concludes this preparation by stating a definition of density-in-phase that emphasizes that it is the number of systems within the extension-in-phase divided by the volume of the extension in the limit as each differential dimension of the volume (${\mathrm{d}q}_i$, ${\mathrm{d}p}_i$ for all $i$) tends to zero. Burbury observes (as Poincaré and Zermelo also had) that, as a mathematical deduction, the extension-in-phase, and therefore the entropy too, are rigorously constant. 

While Burbury and Gibbs agreed that the density-in-phase and entropy should rigorously be constant, Gibbs suggested that ``though Long Periods of Time'' (in the chapter title) the entropy can change. Remarking on this discussion of changing entropy, Burbury \parencite*[p.~255]{burbury_variation_1903} asked, ``if the constancy of $\eta $ [the probability function] depends on the fulfillment of certain conditions, it may be that with lapse of time the conditions will fail, and $\eta $ cease to be constant. But we are not told what conditions, nor that any conditions, will fail.''

Burbury continued to follow Gibbs's exposition coming to the discussion of the addition coloring to water (reviewed above and also discussed by Zermelo, below). Burbury agreed that the mean squared density of coloring matter had been shown to be constant, but in contrast Burbury disagreed with Gibbs's suggestion that an infinite amount of stirring might lead to changes in the mean squared density, and, reflecting his day-job career as a barrister, Burbury \parencite*[p.~256]{burbury_variation_1903} stated, ``I appeal from Professor Gibbs the philosopher of Chapter~XII to Professor Gibbs the mathematician of Chapter I.'' Implicit in this praise of Chapter I is recognition by Burbury that Gibbs had dealt with the \emph{fundamental assumption} by not making any assumptions about atomic velocities.

Burbury proposed a resolution by returning to the definition of density, now of coloring and not number of states, in a differential volume. He based his argumentation on the observation that the medium in Gibbs's illustrative example consists of discrete molecules of finite size. In the limit of differential volume approaching zero, the number of molecules of coloring in that volume will no longer be meaningful. Burbury stated that a new definition of density was needed and proposed that the density about a point $P$ is the quantity of coloring matter in a sphere of finite diameter centered on $P$.\footnote{Burbury had earlier proposed a similar new definition of density in his book on kinetic theory already discussed \parencite[p.~5, \S~9]{burbury_treatise_1899}} Burbury then proposed that a corresponding change to the definition of density-in-phase is possible. In both cases, limiting the volume to a small but non-infinitesimal volume allows the density of coloring molecules, density-in-phase and probability index to change with time. Burbury's use of a sphere of finite diameter in his definition of density is conceptually identical to the small but not infinitesimal volumes, $DV$, used by Gibbs and retained in my synthesis, {\S}~3. What Burbury added was a physical basis for stopping the approach of the volume to zero at a finite dimension: molecular dimensions. Recall that (other than the last chapter) Gibbs had not considered molecules or even dimensions of the systems in his ensembles.

But this new definition of density-in-phase had difficulties that came with it. How is entropy defined in terms of the probability function now? And if this new definition allows the probability function or entropy to change, is that change reversible or irreversible? Those were the questions with which Burbury ended the first paper. It is therefore the point to switch to the response from Henry A. Bumstead.

In that response, Bumstead \parencite*{bumstead_variation_1904} took up the two difficulties presented by Burbury and just discussed: first, long-time stirring and small but non-zero volumes of phase space; and second, the relationship of the new definitions to reversibility and increasing entropy.

Agreement on the first point, or at least an absence of meaningful disagreement, was reached. Bumstead suggested that the long stirring produces smaller and smaller volumes of pure color until the small size of the element of volume becomes a problem. Taking what happens at long time as revealing what happens at short lengths appears to result in clear agreement. In contrast, there does appear to be a lack of agreement as to why short lengths pose a difficulty. Burbury indicated that when the dimensions of an element of volume became comparable to molecular dimensions the density became discontinuous and a new definition of density was needed. Agreeing in part, Bumstead \parencite*[p.~9]{bumstead_variation_1904} stated, ``I nevertheless must agree with Mr. Burbury that{\dots} defining the density by finite elements of volume (or of extension-in-phase) is preferable.'' Bumstead found agreement with this redefinition where Gibbs had written ``if very small differences in phase are neglected,'' as quoted by Bumstead. However, Bumstead \parencite*[p.~9]{bumstead_variation_1904} denied that ``molecular structure in the liquid'' had any bearing and goes on to state, ``but it is because we are unable, owing to the finiteness of our perceptions, {\dots} to recognize very small differences of position in the analogous case of the liquid.''

This brings us back to the issue of how to define entropy in terms of the revised definition of density-in-phase. The arguments about the changes in entropy allowed by Gibbs's theory with the revised definition of density would now apply whether $t'$ comes before $t''$, or in the other order. As Burbury \parencite*[p.~44]{burbury_theory_1904} stated it, ``The remaining difficulty is that the argument of [Gibbs in \emph{Statistical Mechanics}] pp.~150, 151 will work either forwards or backwards. Entropy may, for all that appears, either increase or diminish.'' One possible resolution, that Burbury said was indicated by Gibbs, was to call on experience to interpret the mathematics. In other words, to determine which branch to follow, towards $t'$ or towards $t''$.

At that point Burbury stepped away from considering Gibbs and asked whether it would be possible for a theory of elements with reversible motions to prove that the aggregate motion is irreversible. He concludes that some physical assumption must be made to support the mathematical theorem.

Burbury concluded that correctly redefining Gibbs density-in-phase to allow for discontinuity at molecular dimensions had the consequence of allowing the density-in-phase and entropy to vary with time but without showing whether it increases or decreases or even whether it is monotonic.

\subsection{ Jeans: Continuing Reliance on Boltzmann's Assumptions}

James H.~Jeans would seem to be prepared for commenting on Gibbs \emph{Statistical Mechanics}, both because of his interest and writing on the kinetic theory of gases and because of this training in mathematics. Jeans, among many others, was interested in the disconnect between the number of degrees of freedom of diatomic gases suggested by atomic theory and the equipartition theorem, and by experiment; on the one hand, six degrees of freedom were expected (with some people interpreting emission spectra as indicating the presence of many more degrees) but only five were seen when interpreting measurements (like the speed of sound) in ideal gases. 

In 1903 James H.~Jeans \parencite{jeans_kinetic_1903} published a paper---the response to this paper by Burbury was discussed in the previous subsection---presenting a dynamical approach to the theory of gases. He indicated that the paper was needed because earlier treatment of the kinetic theory of gases rests on a basis (Boltzmann's molecular disorder, or Burbury's assumption A) that, ``can be shown to be neither \textit{a priori} logical, nor \textit{a posteriori} justified by success.'' Jeans's approach was then, ``to follow the motion of a dynamical system starting from an unknown configuration.'' This paper discussed the positions and velocities associated with the degrees of freedom of a gas. Jeans understood dynamical to mean that at each instant the positions and velocities of the constituents of the gas were fully determined by the positions and velocities at a prior instant. What Jeans did not do was to follow the dynamical approach through to the use of the equations of dynamics without reliance on the assumptions about collisions and velocities, as Poincaré and Gibbs had done. Jeans indicated the origins of his theory in earlier work by Maxwell and Boltzmann involving the velocities of gas molecules. 

As previously discussed, Burbury stated that Jeans's theory continued to rely on his Assumption A, discussed above, and Jeans indicated not understanding this critique.

This article from 1903 was followed in 1904 by a textbook that went through a second and third edition. In the preface to that book Jeans stated that he had been motivated to write \emph{The Dynamical Theory of Gases} and provide, ``as exact a mathematical basis as possible.'' \parencite[p.~v]{jeans_dynamical_1904}In this book (now p.~167-8, {\S}~194), we find a brief discussion of Gibbs's example from Chapter~XII of mixing coloring in water; Jeans mixed red and blue ink, and did cite \emph{Elementary Principles in Statistical Mechanics}. There is no other indication of contact with or influence from Gibbs in this textbook.

The second edition of the textbook \parencite{jeans_dynamical_1916} continues the 1904 approach of reliance on kinetic theory. The next successful English-language textbook, Ralph H.~Fowler \emph{Statistical Mechanics}, in contrast, based its exposition on Gibbs's approach to Hamiltonian dynamical systems. In the following prominent English-language textbook in the field, \emph{The Principles in Statistical Mechanics}, Richard C. Tolman \parencite*[p.~ix]{tolman_principles_1979} followed this choice as stated in the first paragraph of the preface, ``It is even hoped that the present book may in some measure meet, from a modern point of view, the same needs for a fundamental exposition as were originally so successfully filled, the classical mechanics, by the treatment given by Gibbs in his \emph{Elementary Principles in Statistical Mechanics}. Throughout this book, although the work of earlier investigators will not be neglected, the deeper point of view and more powerful methods of Gibbs will be taken as ultimately providing the most satisfactory foundation for the development of a modern statistical mechanics.''

Further commentary by Jeans on the work of Gibbs appears in his review of \emph{The Scientific Papers of J. Willard Gibbs} \parencite[p.~144]{jeans_scientific_1907} ; his dramatic statement is worth quoting.

\begin{displayquote}
It is hardly too much to say that notwithstanding his tremendous reputation there has hardly been a scientist in the front rank whose work has been as little studied at first hand, as Professor Gibbs. In his earlier years, the importance of his papers was not understood at all; afterwards it was known only to a few from a first-hand study of the papers. The majority of workers have realized his greatness solely by the far-reaching importance of the results associated with his name: they have regarded him as working in a world of his own creation, as `voyaging through strange seas of thought, alone,'---and they have not read his papers. One cannot of course bring an indictment against the whole scientific world; if Gibbs's papers have been generally regarded as unapproachable, the cause must be looked for in the papers themselves. A large part of this cause is undoubtedly to be found in the somewhat repellant notation and terminology adopted by Gibbs, and in the absence of concrete ideas and illustrations connecting his abstruse mathematical processes with the particular world of thought in which physicists are accustomed to move.
\end{displayquote}

Contrary to our stated expectation, this quote shows that Jeans was not comfortable with the mathematical abstraction in Gibbs's work. This discomfort is likely also connected with his choice not to include any aspect of Gibbs's statistical mechanics in his text book.

\subsection{ Planck: Calculating Entropy}

In a short paper, Max Planck submitted in July 1903 for the \emph{Festschrift} for Boltzmann's 60th birthday and published in February 1904, Planck \parencite*{planck_uber_1904} reviewed the calculation of entropy in Gibbs's and Boltzmann's theories\footnote {Readers may appreciate the irony of the juxtaposition of Gibbs and Boltzmann in a book created to celebrate Boltzmann's birthday. }. Planck noted that Gibbs's definitions have a more general meaning because they do not depend on any special assumptions about the mechanical system being considered and noted the addition of a canonical ensemble. These observations were only brief asides from his focus on comparing the calculation of entropy. On that subject, he satisfied himself that the different methods and formulas produced results that agreed unless the gas contains a mix of different molecules. In that case, Planck stated that Gibbs's grand canonical ensemble needed to be used to get a value for the entropy in agreement with the thermodynamic value. Stated differently, it is necessary to use Gibbs's grand canonical ensemble to resolve Gibbs's paradox. In introducing the grand canonical ensemble in Chapter XV, Gibbs had emphasized the concept of indistinguishable particles. Planck indicated that generally the choice of method for calculating entropy was up to the individual, but the Boltzmann's formula was likely to be easier to use.

\subsection{ Brillouin: Further Discussion of Mixing}

Marcel Brillouin responding to Chapter~XII of Gibbs's book wrote two notes\footnote{These notes by Brillouin follow the second part of the translation arranged by Brillouin of Boltzmann's lectures on gas theory \parencite{boltzmann_cons_1905}; there are two other notes that follow the first part. }. The notes don't constitute a review in themselves but they do have connections with a review by Hadamard. 

In the first note \parencite{brillouin_sur_1905}, Brillouin provided a brief summary of Gibbs's book with a focus on whether it could explain equilibrium: a long-term, steady state. He concluded that it did not, just as the kinetic theory of gases does not account for apparently missing degrees of freedom in the ratio of specific heats. He referred to work by James Jeans\footnote{Presumably, the reference is to \parencite{jeans_partition_1905} or to a section of an earlier article referenced therein \parencite{jeans_distribution_1901}. } suggesting that a resolution of this problem could be found in the interaction of the gas molecules with the ether. Brillouin suggested that such an interaction could also explain irreversible processes.

The second note \parencite{brillouin_sur_1905-1} is about homogeneity and reversibility. In the context of mixing a colored liquid in an uncolored liquid from that chapter, Brillouin introduced the concept of homogeneity referring to the mixing of coloring perceived by a human observer. This was one side of the contrast between human perception and the prediction of the mechanical laws of dynamical systems. Human perception, even with the observational tools available to physicists, had a limit to the small sizes that could be observed. He also indicated that starting with a container of uncolored liquid, there was an apparent organization, a drop, when the coloring was added to the liquid. Then the container was rotated. Brillouin \parencite*[p.~273]{brillouin_sur_1905-1} stated that while rotating the organization once visible but no longer apparent is still real although latent, impossible to observe for all remaining time. He hypothesized that there could be a configuration of water with coloring where rotating the container would cause a latent organization to become apparent. (This suggestion could have some parallel in Poincaré's idea of apparent and latent organization \parencite{poincare_reflexions_1906}.) Defining a volume of phase space surrounding the drop of coloring and tracing the development of the volume is easier than (with time reversed) defining a volume in phase space enclosing widely dispersed coloring that subsequently evolves to surround a drop of coloring. 

The subsequent introduction by Brillouin to the translation into French of Gibbs's book, mentioned above, did not provide commentary on Gibbs's content.

\subsection{ Hadamard: Gibbs, an Introduction for Mathematicians}

A book review by Jacques Hadamard \parencite{hadamard_book_1906} appeared about three and a half years after the above review by Bryan. This span of time better reflects the effort required---Hadamard wrote, ``This book is in no way among those that can be analyzed hastily.''---to study the work and prepare an in-depth review\footnote{All translations from French are provided by the author. }. 

Hadamard \parencite*{hadamard_book_1906} started by observing that ``one of the precious qualities of the work'' is closely associating physics and mathematics. It was, he stated, ``a purely mathematical question'' because it is the application of the calculation of probabilities to ``many systems governed{\dots} by the same equations but with different initial conditions.'' Mathematicians who did not have a sufficient understanding of the kinetic theory of gases needed an initiation; Hadamard proposed for those mathematicians, in fact, most mathematicians, a guide from among their own who spoke their own language, Gibbs.

Since, as noted above, Gibbs used the canonical equations in Hamiltonian form, Gibbs removed all unnecessary elements so that mathematicians could see clearly the path for them to follow. After that observation, Hadamard stated that this development owed much to the work of Poincaré; he specifically refers to Poincaré's \emph{New Methods in Celestial Mechanics}, an application of dynamical systems theory.

Hadamard noted that Brillouin had expressed reservations about the reasoning and results of Boltzmann and Gibbs and presented counterarguments to reasoning presented by each of them. (This is a reference to \parencite{brillouin_sur_1905}, or one of the related notes, which was discussed in {\S}~5.4; Hadamard did specify which note.) Hadamard also suggested that many physicists opposed the ideas of Boltzmann and Gibbs, as Brillouin did. 

Hadamard in contrast expressed, with reserves, support for Boltzmann and Gibbs over the objections of Brillouin. I suspect that Hadamard did not understand that the potential included in the Hamiltonian accounted for the interaction between systems in an ensemble \parencite[sentence split between pages 202 and 203]{hadamard_book_1906}. A better understanding might have led to a different understanding.

Hadamard \parencite*[p.~204]{hadamard_book_1906} indicated that his confidence in Gibbs and Boltzmann depended on two parts. The first was that arguments based on dynamic reversibility do not mean that it is impossible to have final laws that are irreversible. He provided an example to suggest this is possible but it is thin. Second, once there are irreversible laws, there will be a quantity that is monotonically increasing.

Hadamard's choice of where to place his confidence was accompanied by recognition that there were still problems to be resolved. As an example, Hadamard \parencite*[p.~206]{hadamard_book_1906} mentioned that following the trend to infinitesimal extensions in phase would make the application of the law of large numbers in Gibbs's reasoning doubtful or more delicate. Since the averages are being done over many small extensions in phase and not within one small extension, it is unclear that this is a problem that needed to be resolved to produce a satisfactory proof from Gibbs's reasoning in Chapter~XII. It is instead an indication of the need for attention to rigor in working with that chapter.

Hadamard \parencite*{hadamard_book_1906} concluded with his view of the relation between Gibbs's book and Boltzmann's \emph{Lectures on Gas Theory}; the work by Gibbs did not replace the work by Boltzmann or other works that might be undertaken later. Thus, Gibbs's work provided a secure base for further study on the same subject. Hadamard stated that it is indispensable to the successors of Boltzmann since in many cases Gibbs's work alone is the secure and convenient instrument that they need.

\subsection{ Zermelo: Recurrence, Again}

At about the same time as the above review by Hadamard, a book review by Ernst Zermelo \parencite*{zermelo_notizen_1906}\footnote{This is the original article; a translation to English is \parencite[p.~II:571--93]{zermelo_ernst_2013}. Subsequent references are to the English translation. } was published that offered a critique of Gibbs's book and also announced Zermelo's translation of the book. This critique presented the two reservations that Zermelo had mentioned in the foreword to his translation, discussed above. 

The first reservation dealt with the effort to align thermodynamic quantities with statistical mechanical quantities. Gibbs presented an effort to align the equation for energy balance in thermodynamics with the Hamiltonian and other quantities from a dynamical approach. Zermelo \parencite*[p.~II:581]{zermelo_ernst_2013} repeated this alignment from Gibbs of average statistical magnitudes with thermodynamic temperature, energy and entropy, and Zermelo continued, ``But this analogy remains entirely superficial and cannot clarify the mechanical character of the thermodynamical equation.'' Zermelo expanded on this point and there is value in his critique; digging into it however doesn't advance our discussion here.

The second critique is closely related to the dispute between Zermelo and Boltzmann from 1896-97 in connection with recurrence. Gibbs \parencite*[chap.~XII]{gibbs_elementary_1902} started with a proof of Poincaré's recurrence theorem and proceeded cautiously, including using the analogy with coloring mixing in water, to reconcile the theorem with thermodynamic entropy. Zermelo was negative and categorical. Zermelo started his review of Gibbs's reasoning by stating that the recurrence theorem precludes a dynamical system from tending to a final state of equilibrium unless it is already in one; the probability function is independent of time. Zermelo \parencite*[p.~II:585]{zermelo_ernst_2013} contradicted Gibbs's reasoning based on mixing by stating, ``The author tried to find a way here by reinterpretation and analogy.'' Zermelo supported his critique with mathematical considerations. Because the functions are mathematically continuous, the order of evaluating the limits in evaluating the density of the coloring in water does not matter; if the functions were discontinuous, this would need to be incorporated at the outset. This statement was made with mathematical certainty. Zermelo extended the scope of this critique by returning to some of his arguments from the dispute with Boltzmann 10 years earlier. Zermelo concluded with the statement that a statistical theory of heat can only be reconciled with the second law of thermodynamics ``if we resolve to replace as a basis Hamilton's equations of motion by diﬀerential equations already containing the principle of irreversibility in themselves.'' \parencite[p.~II:589]{zermelo_ernst_2013}. That is a very tall order. Compared to the exchange with Boltzmann, here Zermelo placed a stronger constraint on what he thought was required to reconcile a dynamical theory of a gas of many molecules with the second law of thermodynamics. As mentioned above, Burbury \parencite*[p.~48, paragraph 18.(a)]{burbury_theory_1904} demanded ``some physical assumption'' to explain aggregate irreversible motion. Zermelo was more constraining by calling for the motion of the elements to be irreversible.

Zermelo's argument was based on a mathematical understanding of continuity. He however missed a physical consideration affecting mathematical continuity at small dimensions; this is a point that Burbury brought up that was discussed in {\S}~5.2 and that Poincaré included in his discussion of the mixing of coloring. Zermelo did not write about dynamics and reversibility again, so there is no indication whether he found these discussions by Burbury or Poincaré persuasive.

\subsection{ Poincaré: A Shared Approach in Dynamical Systems}

Henri Poincaré had a positive view of Gibbs's work on statistical mechanics and mentioned it in his address to the St. Louis World's Fair \parencite{poincare_letat_1904} also available in translation as \parencite{poincare_present_1906}, but the translation should be viewed with caution because of numerous translation errors.). There, on p.~308, Poincaré had responded to Gibbs's example of mixing discussed in {\S}~3:

\begin{displayquote}
Should a drop of wine fall in a glass of water: whatever the liquid's internal law of motion, we will soon see it colored a pinkish hue and, from that moment, you can shake the container however you want, the wine and water no longer seem able to separate. Thus, here is what would be the model of the irreversible physical phenomenon; hiding a grain of barley in a cup of wheat is easy to do. Finding it again and taking it out of there is nearly impossible. Maxwell and Boltzmann explained all that, but the one who saw most clearly, in a book not read enough because it is a bit difficult to read, is Gibbs, in his \emph{Elementary Principles in Statistical Mechanics}.
\end{displayquote}

A decade earlier Poincaré \parencite*{poincare_mecanisme_1893} had suggested an illustration of irreversibility and mixing repeated here in which a grain of barley is placed in a sack of wheat. The close connection between the two independent statements about mixing grains and liquids is striking. 

And there are connections where the nature is less clear. Gibbs \parencite*{gibbs_elementary_1902} started Chapter~XII with a statement of Poincaré's recurrence theorem from his work on dynamical systems and provided his own proof. Had Gibbs read earlier work by Poincaré in dynamical systems? Is Gibbs's proof an independent rederivation? We don't know. It is further possible that they derived the recurrence theorem at about the same time; such an overlap in their work is supported by the timeline suggested by Klein \parencite*{klein_physics_1990}.

Poincaré \parencite*{poincare_reflexions_1906} wrote an article concerning his thoughts on topics he encountered reading Gibbs's \emph{Statistical Mechanics}. Points suggested by Poincaré's engagement with Gibbs, including for example the need for probability in statistical mechanics and the nature of that probability, are discussed in my preprint \parencite{popp_poincare_2024}.

Although already discussed, it is still worth stating again that Poincaré in his 1890 monograph \emph{The Three-Body Problem and the Equations of Dynamics} and Gibbs in the first three chapters of \emph{Statistical Mechanics} both developed dynamical theories with many bodies that were based on analytical mechanics. Both theories were therefore justified \textit{a priori}, dynamical and without assumptions about the interactions between bodies.

\subsection{ Ehrenfest and Ehrenfest-Afanassjewa: Criticism}

Also in 1906, Tatiana Ehrenfest-Afanassjewa and Paul Ehrenfest \parencite*{ehrenfest-afanassjewa_bemerkung_1906} published their remarks on Gibbs's \emph{Statistical Mechanics}. The remarks started with a brief assessment of the first 10 chapters of his book. Tatiana Ehrenfest-Afanassjewa and Paul Ehrenfest referred to the subject of these chapters as, ``a peculiar mechanical analogy to the reversible processes of thermodynamics.''

Let us pause for a moment and consider this pointed comment on Gibbs's work. Ehrenfest had been a graduate student of Ludwig Boltzmann and his writings over the following years did much to systematize and advocate for the work of Boltzmann. In hindsight this effort by P.~Ehrenfest did much to advance Boltzmann's legacy but at the expense of Gibbs's statistics of canonical ensembles. Arthurs S. Wightman \parencite*[p.~29]{wightman_prescience_1990} quoted a similar dismissal by Ehrenfest and Ehrenfest-Afanassjewa \parencite*[p.~50]{ehrenfest_conceptual_1990} in which they called Gibbs's canonical ensembles, ``an analytical trick.'' Wightman \parencite*[p.~36]{wightman_prescience_1990} seems to be well justified when he later wrote, ``The Ehrenfests{\dots} were the self-chosen prophets of Boltzmann's ideas. In reading their account of Gibbs's work, one has the feeling that they are struggling to be fair but do not quite succeed in concealing their incredulity that anyone would prefer Gibbs's approach to that of Boltzmann.'' Jagdish Mehra \parencite*{mehra_josiah_1998} made a similar statement, ``On the other hand, the Ehrenfests clarified, reformulate[d] and changed many of Boltzmann's concepts in the light of the newer developments. Thus their Boltzmann is an idealized `super-Boltzmann' who had been extremely clear about all points in question, while their Gibbs is a somewhat demoted `sub-Gibbs.' By this interpretation and re-interpretation, their celebrated article caused some mischief, certainly as far as Gibbs was concerned.''

Ehrenfest-Afanassjewa and Ehrenfest \parencite*[p.~89]{ehrenfest-afanassjewa_bemerkung_1906} immediately moved to Gibbs's Chapter~XII, which they described as the starting point of his theory of irreversible phenomena, and stated that the theorem, to which the chapter is dedicated, is unproven because the proof is flawed.

I need to note that there is some ambiguity in the identification of which theorem has the allegedly flawed proof. Gibbs's Chapter~XII starts with a statement and proof of Poincaré's recurrence theorem. In addition to the original proof by Poincaré, Zermelo had also published a proof as part of the dispute between Zermelo and Boltzmann. Criticizing the proof provided by Gibbs without considering the proofs provided by Zermelo and Poincaré seems incomplete. The remainder of Chapter~XII, including the discussion of mixing coloring in water, is an attempt to find a direction to reconcile a dynamical systems view (represented by the recurrence theorem) with human perception of mixing. This discussion seems to be what they were referring to, even if calling it a theorem is inappropriate.

Ehrenfest-Afanassjewa and Ehrenfest's discussion of flaws focused on two points. These are related respectively to the application of the definition of density and to the direction of time.

The first point concerned the application of the definition of density in phase space. Ehrenfest-Afanassjewa and Ehrenfest \parencite*[\S~3--7]{ehrenfest-afanassjewa_bemerkung_1906} looked at two definitions of density $\rho $ and $\Rho$ (read capital rho). For them $\rho $ is defined as the density retained by each particle because the fluid is incompressible and they stated that the mean square of this density does not change over time. In contrast, the other density, $\Rho$ was taken by them as the prevailing density in a parallelepiped after having been divided into many such parallelepipeds of small but non-infinitesimal dimensions. These contrasting definitions are similar to the contrasting ones provided by Gibbs and Burbury. Next Ehrenfest-Afanassjewa and Ehrenfest considered an element of phase space $\Omega '$ at time $t'$ which the flow takes to element $\Omega ''$ at time $t''$. They asked how to align and compare the parallelepipeds and the density in each of them at the two times. They saw a problem (which I don't understand) with how this is handled by Gibbs which they conclude means that the theorem of Chapter~XII is unproven (``daß das Theorem des Kapitels XII unbewiesen ist'').

The second point raised by Tatiana Ehrenfest-Afanassjewa and Paul Ehrenfest \parencite*[p.~96]{ehrenfest-afanassjewa_bemerkung_1906} deals with the direction of time. We first encountered this difficulty in \parencite[p.~44]{burbury_theory_1904} and Ehrenfest-Afanassjewa and Ehrenfest did cite Burbury; the direction of time in Gibbs's discussion of mixing is reversible. In the discussion of Gibbs's consideration and in the discussion by Ehrenfest-Afanassjewa and Ehrenfest of their first point, the discussion takes time $t'$ prior to time $t''$. As Burbury and Ehrenfest-Afanassjewa and Ehrenfest noted, this order is unconstrained and the argument would still be valid with $t''$ prior to $t'$.

At their core, these points can be seen as indicating that Ehrenfest-Afanassjewa and Ehrenfest did not think that Gibbs had resolved Poincaré's fundamental challenge: reconciling a dynamical view that the mean square density of the dye does not vary with our commonplace perception that the color of the wine in the water inextricably and irreversibly becomes a uniform pink.

\subsection{ Lorentz: Teaching}

The works of Hendrik A.~Lorentz discussed in this subsection---and of his graduate student Leonard S.~Ornstein discussed in the following subsection---have an important distinction from the work of the other people reacting to Gibbs's book; whereas the other authors provided critiques or commentary, Lorentz taught and Ornstein applied Gibbs's theory. 

By 1905 Lorentz was aware of Gibbs's book and its basis in mechanics \parencite{lorentz_thermodynamique_1905}. That paper provided a citation, but no specific discussion of the content. In fact, there seems to be no critique or discussion of Gibbs's theory by Lorentz; he just starts teaching and publishes his lecture notes in 1907.

\paragraph{ Abhandlungen, Chapter~XI}

Lorentz had started teaching Gibbs's statistical mechanics within five years of the publication of the book. Lorentz includes the subject in a chapter of his treatise on theoretical physics \parencite[chap.~XI]{lorentz_uber_1907}. Chapter~XI, \emph{On the Second Law of Thermodynamics and Its Relation to Molecular Theories}, is based on his lecture notes and the coverage of Gibbs's work appears in paragraphs~69 to 84 of that chapter. Lorentz states that he is summarizing Gibbs's material. Even so Lorentz uses his notes to introduce both the dynamical system and ensemble concepts with mathematical detail. Lorentz, providing two examples, presents the canonical and microcanonical ensembles and indicates that the microcanonical ensemble corresponds to the ergode considered by Boltzmann, and provides proof that the canonical ensemble leads to a Boltzmann distribution and the equipartition theorem.

More important here than any details or examples is Lorentz's confidence in the value and importance of Gibbs's work that Lorentz expressed by teaching Gibbs's statistical mechanics. Further, that teaching motivated a doctoral student, Ornstein, to look at applications.

\paragraph{ Lectures at Collège de~France}

Lorentz in 1912 gave a lecture series including Gibbs's statistical mechanics \parencite{lorentz_theories_1916}. Lorentz again presented Gibbs's statistical mechanics---this time lecturing at the \emph{Collège de~France} before a more distinguished audience and with greater depth and breadth. 

Lorentz also, to an extent not seen in the \emph{Abhandlungen}, included positive commentary on Gibbs's work. He \parencite*[p.~2]{lorentz_theories_1916} stated, ``Willard Gibbs can be regarded as one of the founders of this new branch of theoretical physics, statistical thermodynamics. After that comes the work of Poincaré, Planck and Einstein.''

The five lectures covered a wide range of topics including subjects from Ornstein's thesis, Hamiltonian mechanics and phase space, ensembles (including a comparison of microcanonical and canonical ensembles, and Einstein's ensembles), fluctuations and critical phenomena, and Brownian motion. The influence of his study of Gibbs's \emph{Statistical Mechanics} is seen throughout these lectures.

\subsection{ Ornstein: Thesis and Applications}

Leonard Ornstein prepared his doctoral thesis, \emph{Applications of Gibbs's Statistical Mechanics}, under the supervision of Lorentz and defended it in March~1908. Leonard Ornstein later applied foundations from his thesis, working with Frits Zernike and independently of Einstein, to building a theory of critical opalescence published in 1914 and later extended. As mentioned in connection with Lorentz, this distinguishes him from other readers of Gibbs considered here.

His thesis \parencite{ornstein_toepassing_1908} had four chapters. I will only provide a very brief summary. The first chapter is a review of the analytical mechanics and ensembles in statistical mechanics. In the second chapter, he considers the velocity and configuration distribution of molecules of finite dimension that cannot overlap (since at short range the force between them is repulsive). This work leads to the second Virial coefficient relating the potential from pairwise molecular interactions to their contribution to the pressure. The third chapter looks at the determination of pressure; a subsequent paper, \parencite{ornstein_determination_1909}, in English, is a response to an objection to this chapter in the thesis. The fourth chapter considers the coexistence of different phases of matter. 

The content of Chapter~4 was later used by Ornstein and Zernike in \parencite{ornstein_accidental_1914} to propose a theory of critical opalescence that offered an alternative to Marian von~Smoluchowski's and A.~Einstein's work and eliminated infinities. By dropping the assumption of independent velocities, Ornstein and Zernike made progress where Einstein had not. \parencite{klein_einstein_1993} Correlations, even over long distances, are needed to explain critical opalescence. Further, Ornstein used grand canonical ensembles, one of two kinds of ensembles that Gibbs added to the one ensemble introduced by Boltzmann.

This application provides direct confirmation of the power and fruitfulness of applying Gibbs's grand canonical ensemble. This application is further noteworthy because it shows the need to include correlations that are contrary to the assumptions of only uncorrelated pairwise interactions built into Boltzmann's collisional (kinetic) theory that Boltzmann justified with his concept of molecular disorder.

\subsection{ Duhem: An Appreciation of Gibbs}

Following the publication of \emph{The Scientific Papers of J Willard Gibbs}, Pierre Duhem published, first as an article in \emph{Bulletin des~Sciences Mathématiques} and then as a separate book, \emph{Josiah-Willard Gibbs} \parencite{duhem_josiah-willard_1908}, a brief summary of the life of Willard Gibbs (drawn from the biographical notice following the preface in volume~1 of Gibbs's papers) and a discussion of his work. Reflecting Duhem's interest in thermodynamics, and his comparatively early adoption (as a graduate student) of Gibbs's approach to thermodynamics, this discussion focuses on that aspect of Gibbs's work.

Duhem's discussion of the scientific content does not go into details and his discussion of \emph{Statistical Mechanics} is also limited.

\subsection{ Kroo: Convergence to Statistical Equilibrium}

In the last of the reactions to Gibbs included here, from early 1911, Jan Kroo published an article arguing a point that he said Gibbs should have proven. \parencite{kroo_uber_1911} This article is hardly worth mentioning except Paul Ehrenfest and Tatiana Ehrenfest-Afanassjewa \parencite[already discussed in {\S}5.9]{ehrenfest_conceptual_1990} quoted from it to critique an example Poincaré used in 1906.

At the time of writing this article, Jan Kroo was a graduate student in mathematics at Göttingen, Germany. He enrolled at Göttingen in September~1906 and submitted his thesis in mathematical physics in October~1912. An article based on the thesis was published in 1913 in \textit{Annalen der Physik}. After receiving his doctorate, Kroo returned to Krakow where he had been born. He was not employed by a university in Poland and worked at a family bank. \parencite{ciesielska_doctoral_2019}

Here our interest is in a statement by Kroo about what Gibbs did not do. With firm conviction, Kroo \parencite*{kroo_uber_1911} stated that a statistical ensemble of systems converges in general with time to a state of statistical equilibrium. This is an implausible assertion. He referred to this as the fundamental theorem of statistical mechanics and also said that Gibbs relied on this theorem without explicitly stating or proving it. Kroo \parencite*[\S~2]{kroo_uber_1911} proposed to provide proofs of existence and stability of this statistical equilibrium. The attempted proof is flawed; the most serious flaw is circular reasoning. In the sentence before equation~8, Kroo claims the existence of the state whose existence he is trying to prove. This is damaging for the proof and for the general credibility of the article.

Note that Ehrenfest and Ehrenfest-Afanassjewa \parencite*[\S~27, pp.~71--2]{ehrenfest_conceptual_1990} expressed a different opinion about \parencite{kroo_uber_1911} that did not consider the proof mentioned above to be flawed. They instead indicated that Kroo's correction was necessary because the ``treatment of Gibbs was incorrect''; they supported this assertion with a cross-reference to their article \parencite{ehrenfest-afanassjewa_bemerkung_1906} considered in {\S}~5.9.

\section{ Conclusion}

Gibbs had produced a theory with solid \textit{a priori} justification and fruitful applications.

Gibbs did attempt an explanation based on mechanics of atoms and molecules for what thermodynamic entropy is and why it monotonically increases. In Chapter~XII, Gibbs approached the subject of entropy and attempted an explanation (that included what Poincaré would later call \emph{coarse graining}). Zermelo was particularly critical of this point. Others in their comments on Chapter~XII saw it, at best, as suggestive and not fully persuasive. Indeed, during the decade before publication, Poincaré had proven that a dynamical system cannot have a state variable that increases monotonically. This remains a curious part of the state of physics.\footnote{It is also curious that physicists have no way to measure entropy. They can measure mass, length and time on most scales; they can also measure temperature; they have no way to measure entropy on any scale. }

Many people, including Poincaré, remarked on the difficulty of understanding \emph{Statistical Mechanics}. Burbury, Poincaré, Lorentz and Ornstein do seem to have managed to understand it. Jeans seems to have not understood it and this is surprising because of his skill with mathematics and in-depth knowledge of kinetic theory. Hadamard may not have fully appreciated that the potential incorporated in the Hamiltonian serves to describe the interactions between the bodies.

The commentary by Burbury, over several of his papers, is important because he understood that what he referred to as a Fundamental Assumption was a defect in Boltzmann's theory that had to be corrected as it was not justified \textit{a priori}; Boltzmann's theory might only be applicable to rarefied gases. He applauded the rigor of Gibbs's application of analytical mechanics. He was disappointed by the treatment of entropy.

Planck reviewed how to use Gibbs's formulas for calculating entropy and concluded that Boltzmann's formula was considerably easier to use. Secondarily, he noted that Gibbs's definitions do not require any special assumptions about the mechanical system and that his grand canonical ensemble is required in some cases.

Ehrenfest and Ehrenfest-Afanassjewa were critical and seemed incredulous that anyone would consider Gibbs's handling of ensembles and statistics. Historians have concluded that they were unfair to Gibbs and aggrandized Boltzmann. Hadamard was critical, but less so than Brillouin and Ehrenfest-Afanassjewa and Ehrenfest. 

Lorentz started teaching Gibbs's \emph{Statistical Mechanics} to both students and professionals; both were published. This was an endorsement backed by his reputation.

Ornstein provided fruitful applications of Gibbs's statistical mechanics including, with Zernike, to critical opalescence, where earlier efforts were not successful because they resulted in infinities for measurable quantities. This showed the fruitfulness of grand canonical ensembles. They produced a successful application of Gibbs's statistical mechanics where Boltzmann's theory was not applicable.

Fruitful applications of a theory built on a foundation with \textit{a priori} justification show the success of the theory in Gibbs's \emph{Statistical Mechanics}.

\textbf{Statements and Declarations}

I have no relevant financial or non-financial interests to disclose.

I did not receive support from any organization for the submitted work.

No datasets were generated or analyzed during the current study

\textbf{Credit}

I am the sole author and there are no roles to be provided for other authors.

\printbibliography

\end{document}